\documentclass[reprint,amsmath,amssymb,aps]{revtex4-1}
\usepackage[dvipdfmx]{graphicx}
\usepackage{dcolumn}
\usepackage{bm}
\usepackage{amsmath,amssymb}

\usepackage{color}
\begin{document}

\title{Anisotropic Tensor Renormalization Group}
\author{Daiki Adachi$^1$}
\author{Tsuyoshi Okubo$^1$}
\author{Synge Todo$^{1,2}$}
\affiliation{
  $^1$Department of Physics, University of Tokyo, Tokyo, 113-0033, Japan\\
  $^2$Institute for Solid State Physics, University of Tokyo, Kashiwa, 277-8581, Japan
}

\begin{abstract}
 We propose a new tensor renormalization group algorithm, Anisotropic Tensor Renormalization Group (ATRG), for lattice models in arbitrary dimensions.
The proposed method shares the same versatility with the Higher-Order Tensor Renormalization Group (HOTRG) algorithm, \textit{i.e.,} it preserves the lattice topology after the renormalization.
In comparison with HOTRG, both of the computation cost and the memory footprint of our method are drastically reduced, especially in higher dimensions, by renormalizing tensors in an anisotropic way after the singular value decomposition.
We demonstrate the ability of ATRG for the square lattice and the simple cubic lattice Ising models.
Although the accuracy of the present method degrades when compared with HOTRG of the same bond dimension, the accuracy with fixed computation time is improved greatly due to the drastic reduction of the computation cost.
\end{abstract}

\maketitle

\section{Introduction}
\label{sec:introduction}

Understanding critical phenomena observed universally in many-body systems is one of the central topics in statistical physics.
Due to the difficulties in solving the many-body systems analytically, however, we often have to rely on numerical methods, such as the Monte Carlo method, in the investigation of such complex systems.
In recent years, on the other hand, new alternatives, the real-space renormalization group methods, have become popular more and more and many researches have been conducted using these approaches.
The {\em Tensor Renormalization Group} (TRG), proposed by Levin and Nave in 2007, calculates approximately the contraction of a tensor network by using the singular value decomposition (SVD)~\cite{LevinN2007}.
By representing the partition function of the Ising models by a tensor network, the free energy for the honeycomb or square lattice Ising models is calculated for huge system sizes, which can be regarded virtually as in the thermodynamic limit.

Despite the success of TRG for two-dimensional systems, it is difficult to extend TRG to higher dimensions, because the framework of TRG is tightly related to the two-dimensional lattice topology.
In 2012, Xie {\em et al}. proposed the {\em Higher-Order Tensor Renormalization Group} (HOTRG) method, which performs the real-space renormalization of tensor networks based on the higher-order singular value decomposition~\cite{XieCQZYX2012}.
Unlike TRG, HOTRG can be applied to arbitrary dimensions.
Although the computation cost of HOTRG is higher than TRG, its accuracy is higher when we compare these two methods with the same bond dimension $\chi$, which is the maximum number of the tensor indices.
A lot of methods inspired by TRG and HOTRG have been proposed~\cite{GuW2009, XieJCWX2009, EvenblyV2015, Evenbly2017, BalMHV2017, YangGW2017, HauruDM2018, Harada2018, MoritaIZK2018, WangQZ2014, ZhaoXXI2016, Evenbly2018, Ferris2013}, and they have been used for examining classical and quantum systems not only in the condensed matter physics but also in the particle physics~\cite{LiGZRGS2010, ZhaoXCWCX2010, ChenQCWZNX2011, DittrichE2012, JiangWX2008, EvenblyV2016, Meurice2013, WangXCBX2014, YuXMLDZQCX2014, UedaON2014, GenzorGN2016, YangLZXM2016, KawauchiT2016, SakaiTY2017, YoshimuraKNTS2018, AkiyamaKYY2019}.

Although HOTRG can be applied to higher-dimensional systems in principle, its computation cost in the $d$-dimensional lattice model increases quite rapidly as $O(\chi^{4d-1})$, where $d$ is the dimension of the lattice, as a function of bond dimension $\chi$.
Thus, it is totally impractical to apply HOTRG for large bond dimensions and/or high spatial dimensions.
New algorithms that can perform the real-space renormalization in general dimensions with a reasonable cost are strongly demanded.

In the present paper, we propose a new TRG method, referred to as {\em Anisotropic Tensor Renormalization Group} (ATRG), in which the computational complexity for systems on hypercubic lattices is drastically reduced by renormalizing tensors in an anisotropic way after the tensor decomposition.
The computation cost of ATRG scales as $O(\chi^{2d+1})$, which is much lower than that of the conventional HOTRG, $O(\chi^{4d-1})$, and the memory footprint of ATRG scales as $O(\chi^{d+1})$, which is also much smaller than that of HOTRG, $O(\chi^{2d})$.

The present paper is organized as follows.
We describe the two- and three-dimensional ATRG algorithms in Sec.~\ref{sec:algorithms}. ATRG can be applied to dimensions higher than three in a straightforward way.
In order to further accelerate the algorithm and reduce the memory footprint, especially in higher dimensions, we introduce several additional techniques in Sec.~\ref{sec:techniques}.
In Sec.~\ref{sec:benchmarks}, we calculate the free energies of the square lattice and simple cubic lattice Ising models at their critical points as demonstrations of ATRG.
Finally, we discuss the advantages of out method over TRG and HOTRG in Sec.~\ref{sec:summary}.

\section{Algorithms}
\label{sec:algorithms}

\begin{figure}[tbp]
  \begin{center}
  \includegraphics[clip, width=7cm]{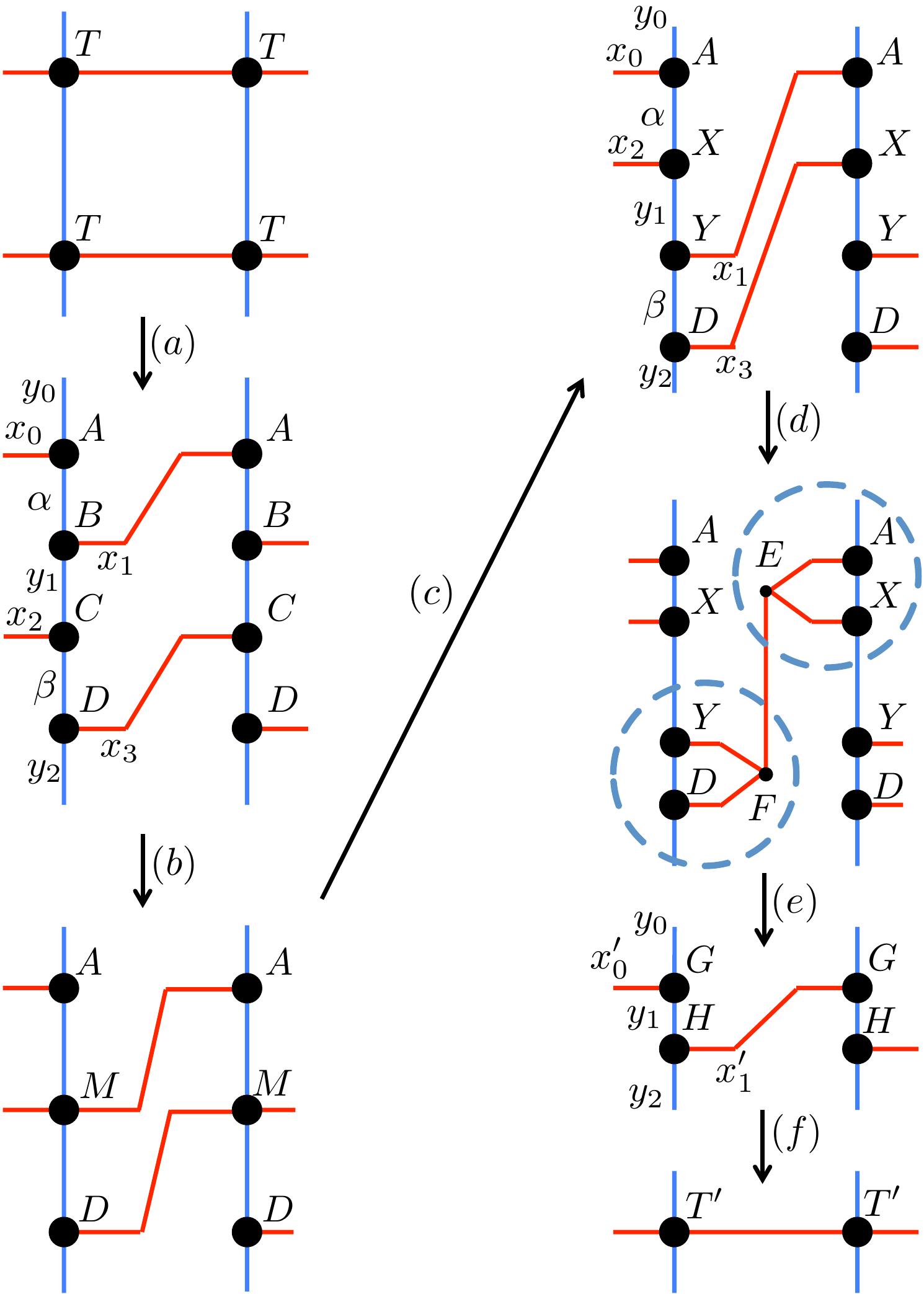}
  \end{center}
  \caption{Renormalization step of ATRG in $y$ direction for the two-dimensional square lattice model.}
  \label{atrg_2d}
\end{figure}

Let us begin with describing the ATRG algorithm for the two-dimensional square lattice network. Hereafter, $U_{\{X\}}$, $S_{\{X\}}$, and $V_{\{X\}}$ represent respectively the left isometry, the singular value matrix, the right isometry of SVD of matrix $X$, i.e., $X=U_{\{X\}} S_{\{X\}} V_{\{X\}}^{t}$ or $X_{ij} = \sum_\alpha S_{\{X\}\alpha\alpha} U_{\{X\}i\alpha} V_{\{X\}j\alpha}$.
In ATRG, we renormalize two neighboring tensors in the horizontal ($x$) and the vertical ($y$) directions alternately as in the same way as HOTRG.
The outline of ATRG renormalization step along $y$ direction is shown in Fig.~\ref{atrg_2d}.
%{{{ ATRG algorithm in 2D
%}}}

First, four-leg tensor $T$ with bond dimension $\chi$ is decomposed by partial SVD, and the bond dimension of the singular value matrix is truncated by $\chi$:
%{{{ equation (a)
\begin{align}
  T_{y_0 y_1 x_0 x_1} \approx \sum_{\alpha=1}^\chi S_{\{T\} \alpha \alpha} U_{\{T\} y_0 x_0 \alpha} V_{\{T\} y_1 x_1 \alpha }. \label{T}
\end{align}
%}}}
Then, we define four tensors, $A$, $B$, $C$, and $D$, as
%{{{ A ~ D definitions
\begin{align}
  A_{y_0 x_0 \alpha} &= U_{\{T\} y_0 x_0 \alpha}, \label{A} \\
  B_{y_1 x_1 \alpha} &= S_{\{T\} \alpha \alpha} V_{\{T\} y_1 x_1 \alpha}, \label{B} \\
  C_{y_1 x_2 \beta} &= S_{\{T\} \beta \beta} U_{\{T\} y_1 x_2 \beta} , \label{C} \\
  D_{y_2 x_3 \beta} &= V_{\{T\} y_2 x_3 \beta} \label{D}
\end{align}
%}}}
[step~(a) in Fig.~\ref{atrg_2d}]. Note that the accuracy of the final free energy may depend on how to separate the singular value matrix in the above step. Indeed, we observe that by including the singular matrix $S$ in $B$ and $C$, the error of the final free energy is minimized. Such a construction gives us a better free energy than the equal weight decomposition, $\sqrt{S}$, of the singular matrix into $A$ and $B$ (or $C$ and $D$). 
Indeed, it is easy to see if we do not introduce truncations in step~(a), our splitting of singular matrix $S$, together with step~(b) and (c) below, becomes identical to direct partial SVD of $TT = ABCD$.
In this sense, the present splitting gives the best local approximation, and we think it remains optimal even if there exist truncations in step~(a).

Next, by using partial SVD, we swap the bond of $B$ and $C$~[step~(b) and (c) in Fig.~\ref{atrg_2d}]. In order to swap the $x_1$ bond of $B$ and $x_2$ bond of $C$, we define tensor $M$ as
%{{{ M definitions
\begin{equation}
  M_{\alpha \beta x_1 x_2} = \sum_{y_1} B_{y_1 x_1 \alpha} C_{y_1 x_2 \beta}, \label{M1}  
\end{equation}
%}}}
and, by partial SVD of $M$ and truncating the singular values to $\chi$, we define new $X$ and $Y$ as 
%{{{ equation (b) and (c)
\begin{align}
  M_{\alpha \beta x_1 x_2} &\approx \sum_{y_1}^\chi S_{\{M\} y_1 y_1} U_{\{M\} \alpha x_2 y_1} V_{\{M\} \beta x_1 y_1}, \label{M2} \\
  X_{\alpha x_2 y_1} &= \sqrt{S_{\{M\} y_1 y_1}} U_{\{M\} \alpha x_2 y_1}, \label{B2} \\
  Y_{\beta x_1 y_1} &= \sqrt{S_{\{M\} y_1 y_1}} V_{\{M\} \beta x_1 y_1}. \label{C2}
\end{align}
%}}}

Then, we renormalize the horizontal two bonds into one by using `squeezers' $E$ and $F$ [step~(d) and (e) in Fig.~\ref{atrg_2d}]. We call them as squeezers since they are not necessarily isometries unlike conventional HOTRG. By applying squeezer $E$ ($F$) to $A$ and $X$ ($Y$ and $D$), we obtain new tensor $G$ ($H$) as
%{{{ definitions of H and G
\begin{align}
  &G_{y_0 y_1 x_0'} = \sum_{\alpha, x_0, x_2} A_{y_0 x_0 \alpha} X_{\alpha x_2 y_1} E_{x_0 x_2 x_0'}, \label{G} \\
  &H_{y_1 y_2 x_1'} = \sum_{\beta, x_1, x_3} D_{y_2 x_3 \beta} Y_{\beta x_1 y_1} F_{x_1 x_3 x_1'}. \label{H}
\end{align}
%}}}
Finally, a new renormalized tensor, $T^{\prime}$, is made from the product of $G$
and $H$ as 
%{{{ equation (f)
\begin{align}
  T^{\prime}_{y_0 y_2 x_0' x_1'} = \sum_{y_1} G_{y_0 y_1 x_0'} H_{y_1 y_2 x_1'} \label{T2}
\end{align}
%}}}
[step~(f) in Fig.~\ref{atrg_2d}], which will be used as an input to the next renormalization step in $x$ direction.

It should be noted that the explicit form of the squeezers is not needed for calculating the free energy. One can obtain $G$ and $H$ directly by partial SVD as
\begin{align}
  Q_{y_0 y_1 y_2 y_3} &= \sum_{x_1 x_3 \alpha \beta} Y_{\beta y_1 x_1} D_{y_2 x_3 \beta} A_{y_0 x_1 \alpha} X_{\alpha x_3 y_3} \nonumber \\
                       &\approx \sum_{x_1'}^\chi S_{\{Q\} x_1' x_1'} U_{\{Q\} y_1 y_2 x_1'} V_{\{Q\} y_0 y_3 x_1'}, \label{C3} \\
  H_{y_1 y_2 x_1'} &= \sqrt{S_{\{Q\} x_1' x_1'}} U_{\{Q\} y_1 y_2 x_1'}, \label{H2}\\
  G_{y_0 y_3 x_1'} &= \sqrt{S_{\{Q\} x_1' x_1'}} V_{\{Q\} y_0 y_3 x_1'}. \label{G2}
\end{align}
The explicit form of the squeezers is needed for calculating other physical quantities, \textit{e.g.,} the energy and the magnetization.
We will discuss the method to calculate squeezers in Sec.~\ref{sec:techniques}.

One of the most significant differences between ATRG and HOTRG is that in ATRG, before applying the squeezers $E$ and $F$, we construct $X$ and $Y$ from a low-rank approximation of $M$ [step (b) and (c) in Fig.~\ref{atrg_2d}], where the singular value matrix is truncated to $\chi$.
ATRG thus involves an additional approximation compared with HOTRG.
Due to this truncation, unlike TRG and HOTRG, ATRG does not sustain axisymmetry of the networks any more.
It is interesting to see that despite the lack of the symmetry, ATRG in two dimensions still has the corner double line (CDL) tensor as a fixed-point tensor as TRG and HOTRG~\cite{GuW2009}.

The computation cost of the whole ATRG renormalization procedure described above is dominated by SVD, that is, it will be proportional to $\chi^6$, if we use full SVD.
The cost can be reduced to $O(\chi^5)$ by adopting partial SVD using the Arnoldi method or other techniques.
The computation cost of ATRG is thus the same as TRG with partial/randomized SVD or projective truncation method~\cite{MoritaIZK2018, NakamuraOT2019} in the two-dimensional case.
The memory footprint is $O(\chi^4)$, which is needed to store the intermediate tensor $M$. As we will see in Sec.~\ref{sec:techniques}, however, the memory footprint can be reduced to $O(\chi^3)$.

\begin{figure}[tbp]
  \begin{center}
  \includegraphics[clip, width=7cm]{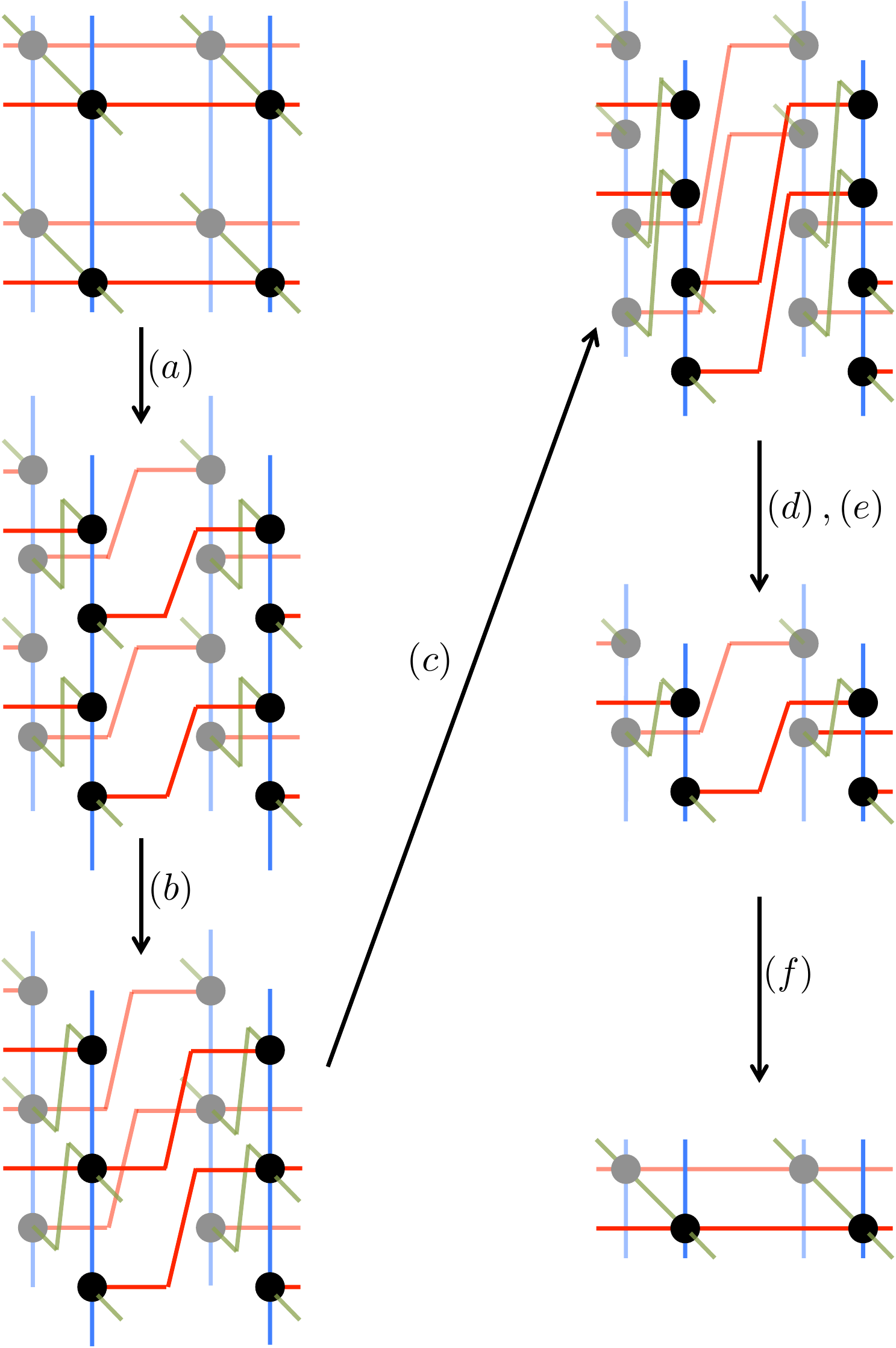}
  \end{center}
  \caption{
    Renormalization step of ATRG in $z$ direction for the three-dimensional simple cubic lattice model.
  }
  \label{atrg_3d}
\end{figure}

One can easily generalize ATRG to systems in dimensions higher than two.
In Fig.~\ref{atrg_3d}, we illustrate the renormalization step of ATRG in three dimensions.
For $d$-dimensional hypercubic lattices, ATRG can renormalize tensors with computation cost of $O(\chi^{2d+1})$ and memory footprint of $O(\chi^{d+1})$.

%{{{ ATRG algorithm in 3D 
%}}}

\section{Techniques}
\label{sec:techniques}

In this section, we discuss several techniques in detail to achieve the optimal cost in ATRG.

\begin{figure}[tbp]
  \begin{center}
  \includegraphics[clip, width=7cm]{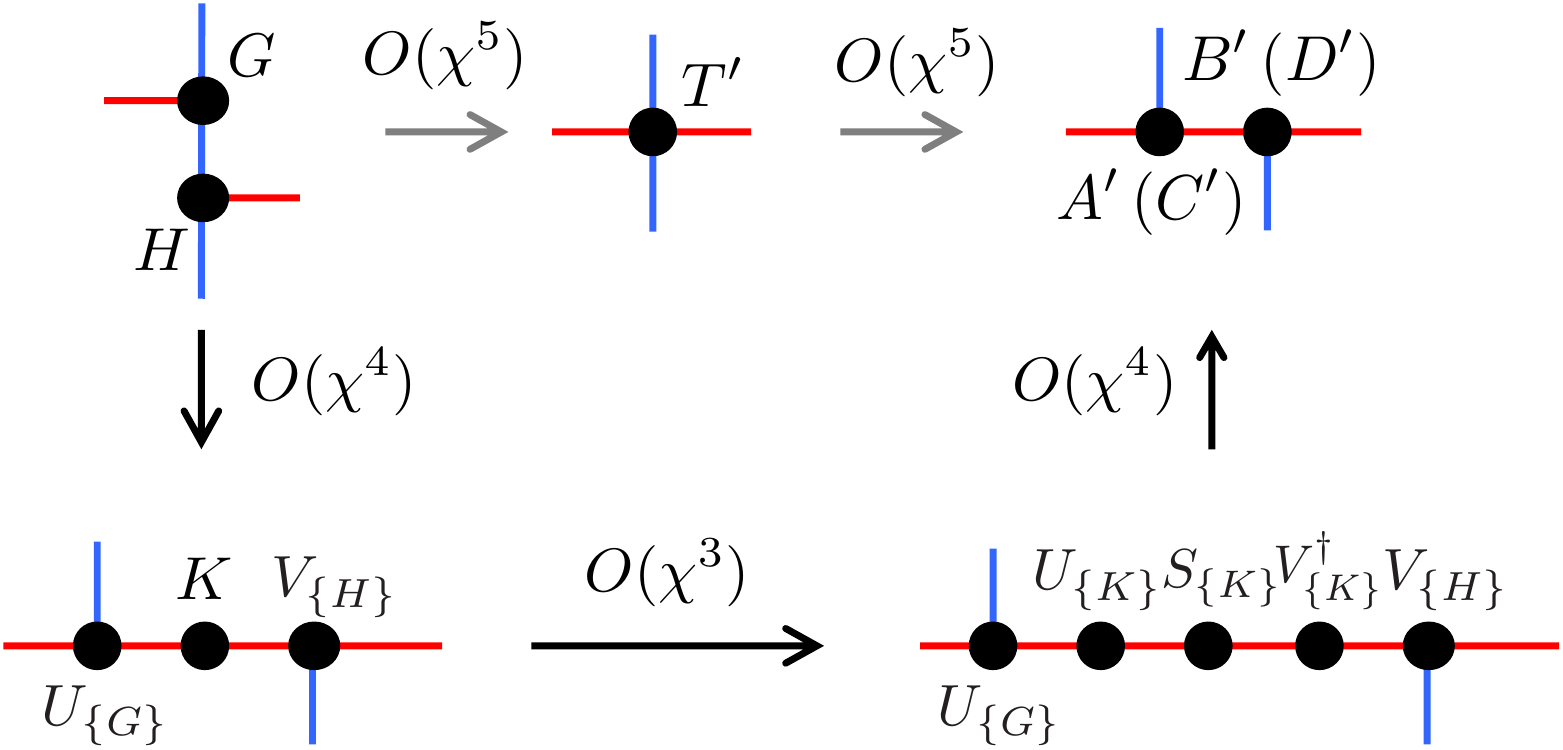}
  \end{center}
  \caption{Reduction of SVD computation cost for the two-dimensional square lattice model. Here, $K=S_{\{G\}} V_{\{G\}}^{t} U_{\{H\}} S_{\{H\}}$.}
  \label{shortcut_2d}
\end{figure}

First, in step~(a) of Fig.~\ref{atrg_2d}, there remains freedom in choosing the bond geometry of $A$, $B$, etc.
If bond geometry of $A^{\prime}$ and $B^{\prime}$ (or $C^{\prime}$ and $D^{\prime}$) in the following renormalization step so as to match the geometry of $G$ and $H$ in the previous step, respectively, we can avoid SVD of $T^{\prime}$, which introduces additional truncation errors, and continue renormalization procedure by using SVDs for $G$ and $H$, which have no truncation error.
The modified step is illustrated in Fig.~\ref{shortcut_2d}.
Note that explicit SVD of $T^{\prime}$ requires the computation cost of $O(\chi^5)$, while in the modified procedure the cost is reduced to $O(\chi^4)$ that includes SVDs of $G$ and $H$, and also of the intermediate two-bond tensor, $K$, defined as $K=S_{\{G\}} V_{\{G\}}^{t} U_{\{H\}} S_{\{H\}}$.

%{{{ shortcut picture
%}}}

\begin{figure}[tbp]
  \begin{center}
  \includegraphics[clip, width=4.5cm]{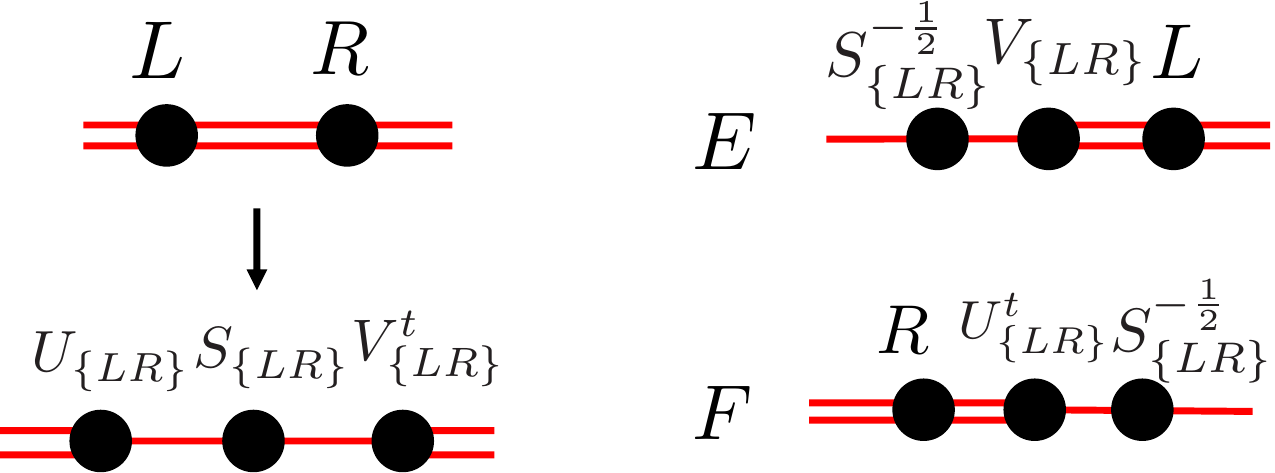}
  \end{center}
  \caption{Preparation of squeezers $E$ and $F$ in step (d) in Fig.~\ref{atrg_2d}. These squeezers $E$ and $F$ are chosen so as to minimize $\|LR - LFER\|^2$.}
  \label{make_EF}
\end{figure}

Next, we explain how to obtain squeezers $E$ and $F$ in step~(d) in Fig.~\ref{atrg_2d}.
In this step, as we explained, we want to prepare squeezers $E$ and $F$ that satisfy Eqs.~\eqref{G}-\eqref{G2}.
In general, such squeezers should minimize $\|LR - LFER\|^2$, where $L=YD$, $R=AX$, and $\|\cdot\|$ denotes the Frobenius norm, and thus they can be obtained from the results of partial SVD of $LR$ as
\begin{align}
  E = S_{\{LR\}}^{-\frac{1}{2}}V_{\{LR\}}L \label{E} \\
  F = RU_{\{LR\}}^{t}S_{\{LR\}}^{-\frac{1}{2}} \label{F}
\end{align}
(Fig.~\ref{make_EF})~\cite{CorbozRT2014}.

%{{{ make_EF
%}}}

%{{{ truncation technique
\begin{figure}[tbp]
  \begin{center}
  \includegraphics[clip, width=8cm]{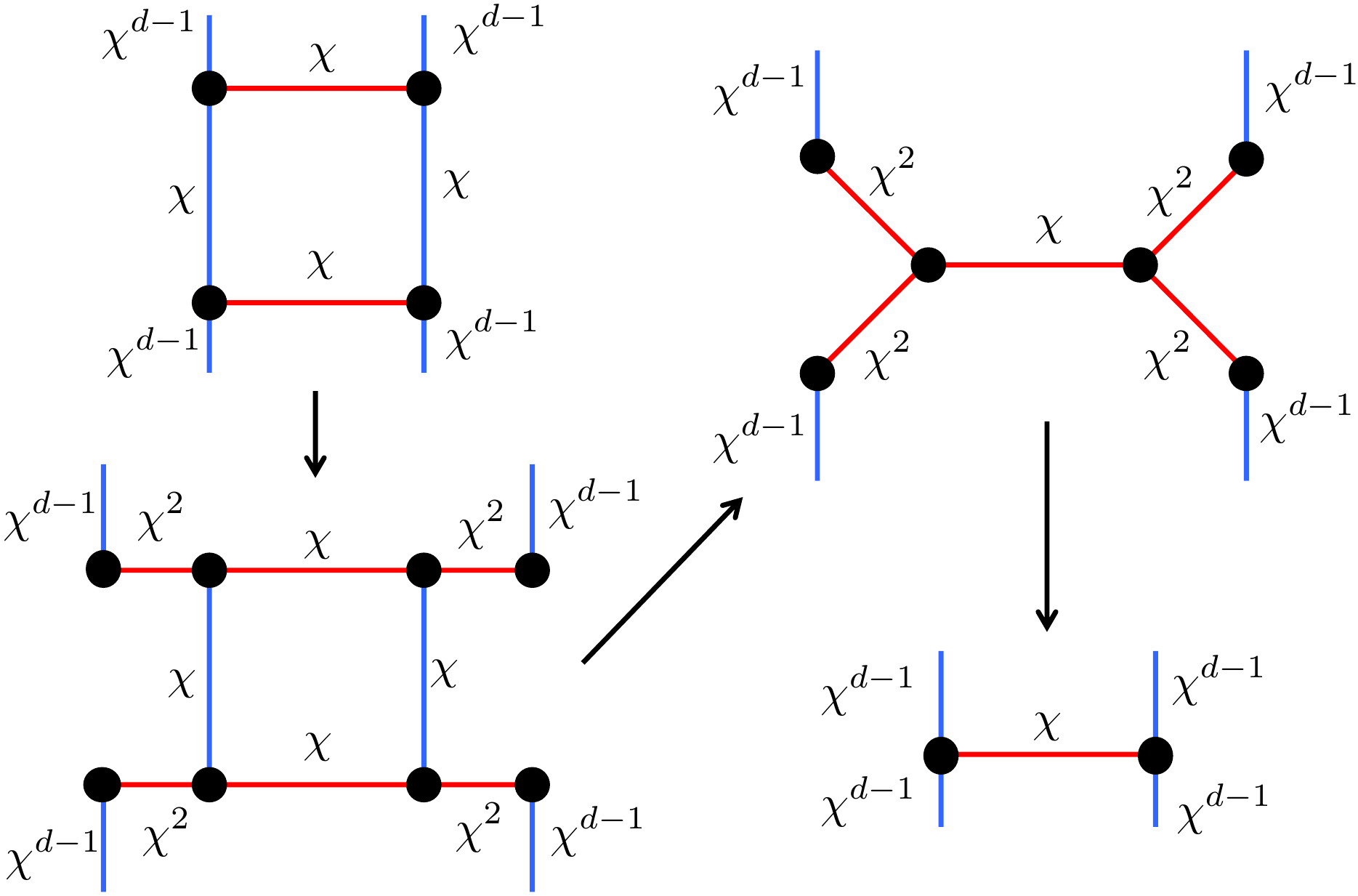}
  \end{center}
  \caption{
    Reduction of SVD cost in step (d) and (e) in Fig.~\ref{atrg_3d} for $d>3$.
  }
  \label{trunc_tec}
\end{figure}
%}}}

In the above procedure, the cost for obtaining the squeezers is naively expected to be the same as that of partial SVD, i.e., $O(\chi^{2d+1})$.
However, it can be reduced to $O(\chi^{\max(d+3, 7)})$ as shown in Fig.~\ref{trunc_tec}, where $A$, $X$, $Y$, and $D$ in Fig.~\ref{atrg_2d} are represented as three-bond tensors with bond dimensions $\chi^{d-1}, \chi, \chi$.
In Fig.~\ref{trunc_tec}, before making squeezers, SVD or QR decomposition is performed for each tensor.
By introducing these preprocess, the cost of SVD for making squeezers in step~(d) and (e) in Fig.~\ref{atrg_3d} is reduced to $O(\chi^{\max(d+3, 7)})$, which means the cost of SVDs for making squeezers becomes subleading for $d>3$.

\begin{figure}[tbp]
  \begin{center}
    \includegraphics[clip, width=7cm]{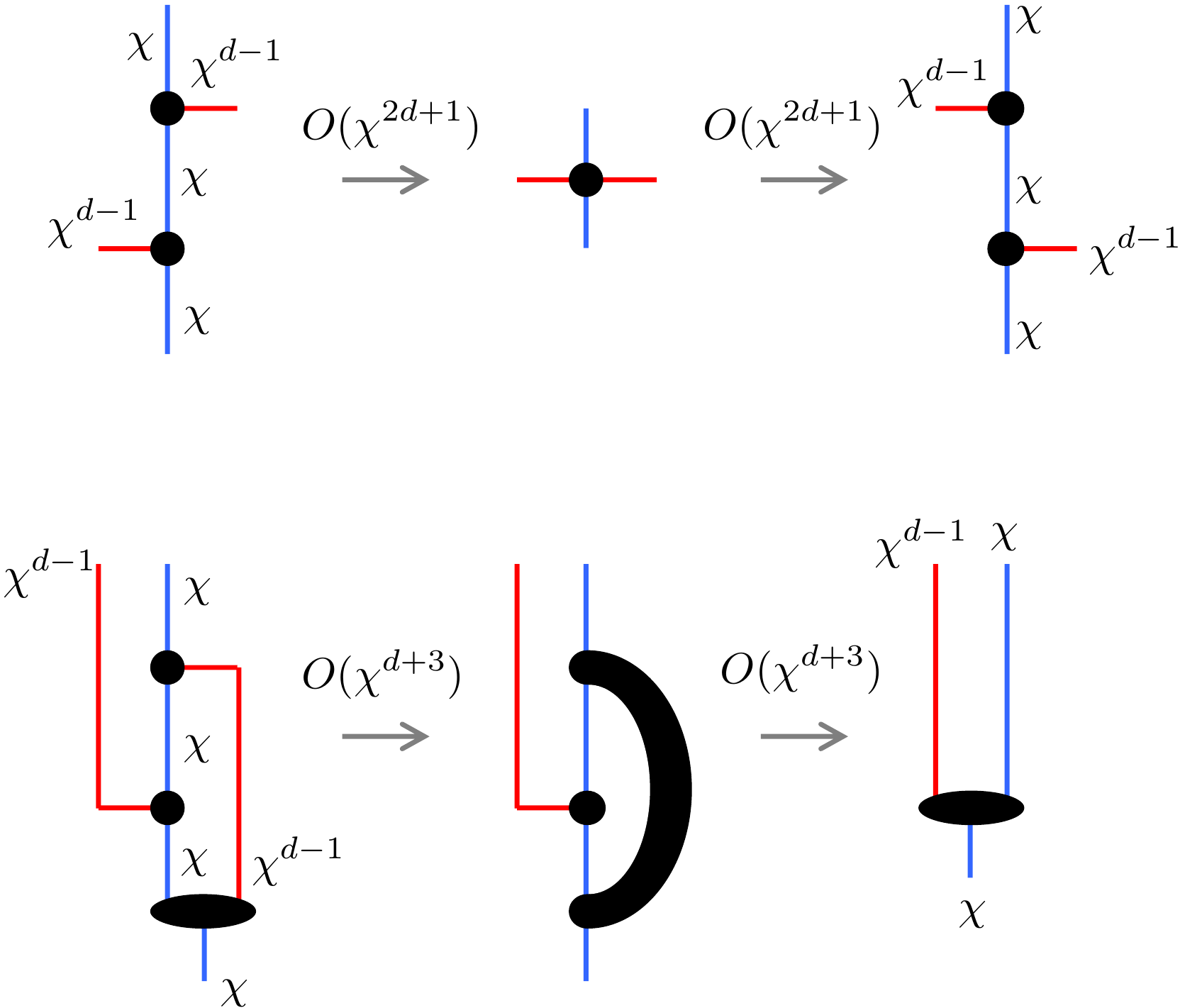}
  \end{center}
  \caption{
    Factorization of matrix-vector multiplication in SVD.
    The upper row is the original procedure [step (b) and (c)] presented in Fig.~\ref{atrg_2d}.
    The memory footprint of the intermediate tensor is $O(\chi^{2d})$.
    In the lower row, the matrix-vector multiplication in partial SVD is factorized in two successive tensor-vector multiplications.
    This procedure reduces the memory footprint from $O(\chi^{2d})$ to $O(\chi^{d+1})$ and also reduces the computation cost from $O(\chi^{2d+1})$ to $O(\chi^{d+3})$.
  }
  \label{memory_tec}
\end{figure}

Finally, we discuss the computation cost for swapping bonds [step~(c)]. The cost for this procedure can be reduced to $O(\chi^{d+3})$.
In addition, the memory footprint can also be reduced from $O(\chi^{2d})$ to $O(\chi^{d+1})$.
In Fig.~\ref{memory_tec}, we show the procedure to reduce memory in step~(b) and (c) in Fig.~\ref{atrg_2d} in $d$ dimensions.
The upper sequence in Fig.~\ref{memory_tec} is the original procedure shown in Fig.~\ref{atrg_2d}, where
the memory footprint of the intermediate tensor is $O(\chi^{2d})$.
In partial SVD based on the Arnoldi method, the most time-consuming part is the matrix-vector multiplication.
The matrix-vector multiplication can be factorized into two successive tensor-vector multiplications as shown in the lower row of Fig.~\ref{atrg_2d}.
By this way, generating a large intermediate tensor can be avoided.
This procedure requires only the memory footprint of $O(\chi^{d+1})$ (to store $\chi$ vectors of length $\chi^d$) and the computation cost of $O(\chi^{d+3})$.

%{{{ memory_save technique
%}}}

\section{Benchmarks}
\label{sec:benchmarks}

In order to demonstrate the efficiency of ATRG, we apply ATRG to the two- and three-dimensional Ising models.
In both cases, each axis ($x$ and $y$, or $x$, $y$, $z$) is renormalized $15$ times alternately, \text{i.e.,} the system contains $(2^{15})^d$ spins.

\begin{figure}[tbp]
  \centering
  \includegraphics[width=7cm]{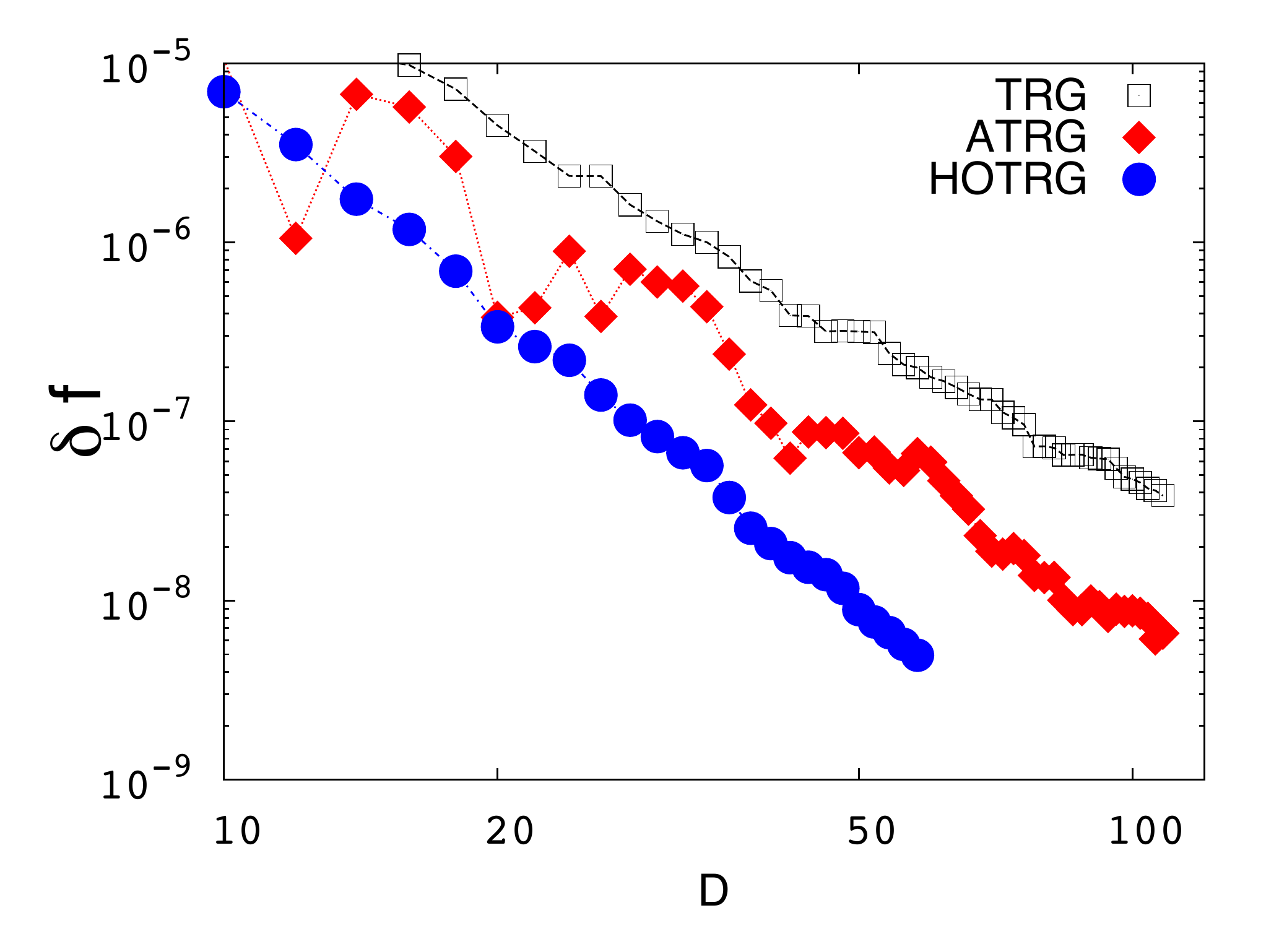}
  \vspace{15pt}
  \caption{
    Absolute error of the free energy density of the two-dimensional Ising model at $T=T_{\rm c}$ as a function of bond dimension $\chi$ calculated by TRG (black squares), HOTRG (blue circles), and ATRG (red diamonds). Even though ATRG has the same computation cost as TRG, it produces more accurate results than TRG.
  }
  \label{Ising2D_1}
%\end{figure}
%\begin{figure}[tbp]
  \centering
  \includegraphics[width=7cm]{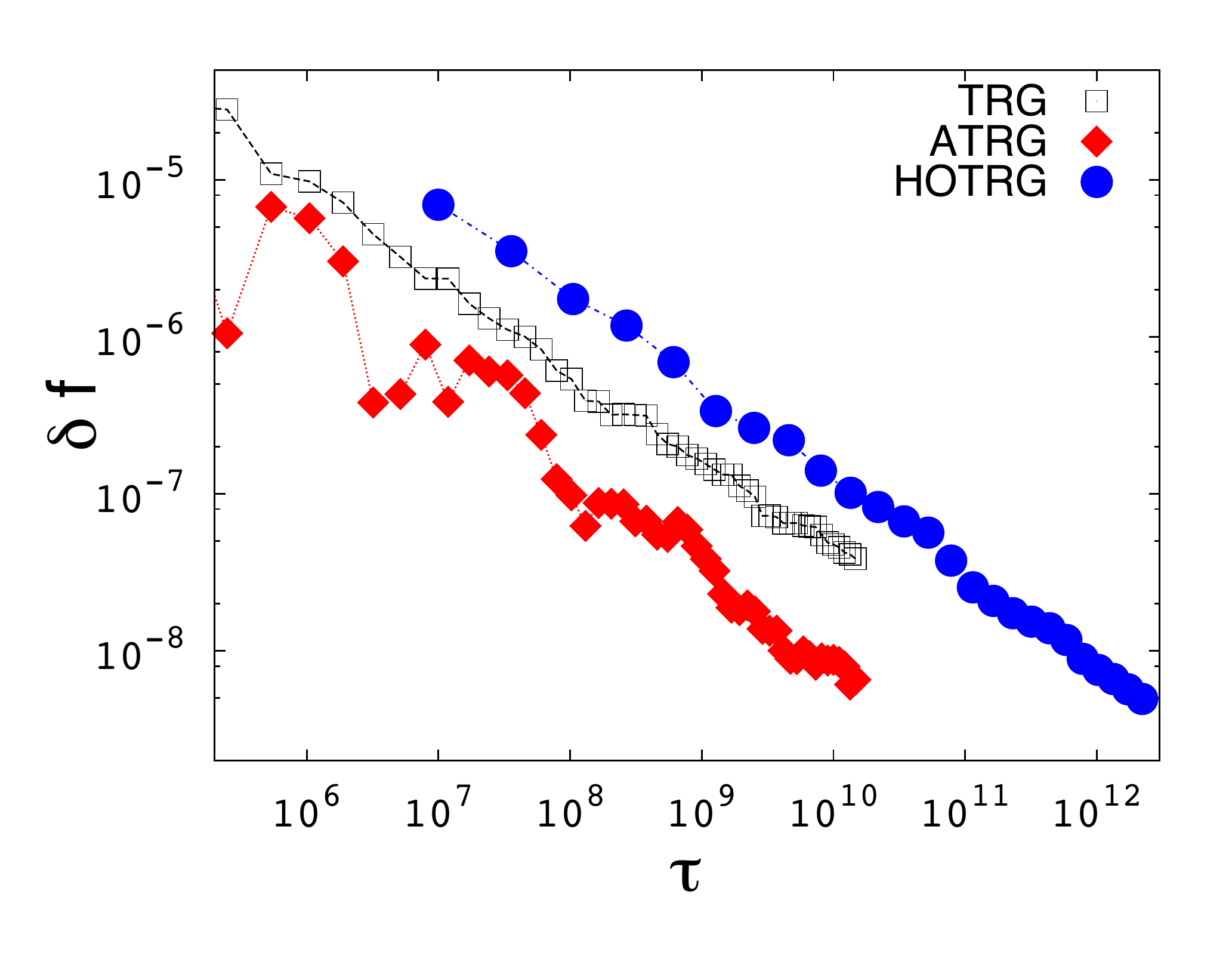}
  \vspace{15pt}
  \caption{
    Absolute error of the free energy density of the two-dimensional Ising model at $T=T_{\rm c}$ as a function of leading-order computation time $\tau$ [Eq.~\eqref{tau_2d}] calculated by TRG (black squares), HOTRG (blue circles), and ATRG (red diamonds). ATRG achieves the most accurate results among the three methods with fixed computation time.
  }
  \label{Ising2D_2}
\end{figure}

First, we discuss the two-dimensional Ising model on a square lattice.
The free energy density calculated at the critical temperature, $T = T_{\rm c} = 2 / \text{log}(1 + \sqrt{2})$~\cite{Onsager1944}, is shown in Figs.~\ref{Ising2D_1} and \ref{Ising2D_2}.
In Fig.~\ref{Ising2D_1}, we plot the absolute error of the free energy density as a function of bond dimension $\chi$.
We calculate up to $\chi=108$ ($\chi=58$) for TRG and ATRG (HOTRG).
For all $\chi$, the result of ATRG is between those of TRG and HOTRG.
Note that ATRG has the same computation cost as TRG, while the cost of HOTRG is higher.
In order to investigate the performance of ATRG more precisely, we compare the $\chi$ dependence of the error with fixed computation time. For that purpose, we introduce the leading-order computation time $\tau$ (dimensionless quantity) defined as
\begin{align}
  \tau = \begin{cases}
    \chi^5 & \text{for TRG and ATRG} \\
    \chi^7 & \text{for HOTRG}
  \end{cases}
  \label{tau_2d}
\end{align}
based on their leading computation cost.
In Fig.~\ref{Ising2D_2}, we plot the absolute error of the free energy density as a function of $\tau$ for the three methods.
With fixed $\tau$, ATRG has the smallest error among the three methods.
In ATRG, partial SVD is the most expensive operation, while it is the contraction in HOTRG.
In practice, the partial SVD takes much longer time than the contraction, even when their computation costs are in the same order.
Thus, the actual performance difference between ATRG and HOTRG is smaller than Fig.~\ref{Ising2D_2}, though ATRG becomes more and more advantageous for larger $\chi$ due to the difference in order in computation cost [Eq.~\eqref{tau_2d}].

%{{{ RESULTS OF 2D ISING
%}}}

It should be mentioned that the present method suffers from larger and nonmonotonic fluctuations in the convergence of the error.
This observed behavior is probably related to the two independent truncations in the ATRG renormalization procedure.
Because ATRG optimizes only the local tensors in each truncation, increasing $\chi$ does not necessarily improve the accuracy of the free energy, which is determined by the global tensor network.
The nonmonotonic convergence in ATRG should be subjected to further investigation.

\begin{figure}[tbp]
  \centering
  \includegraphics[width=7cm]{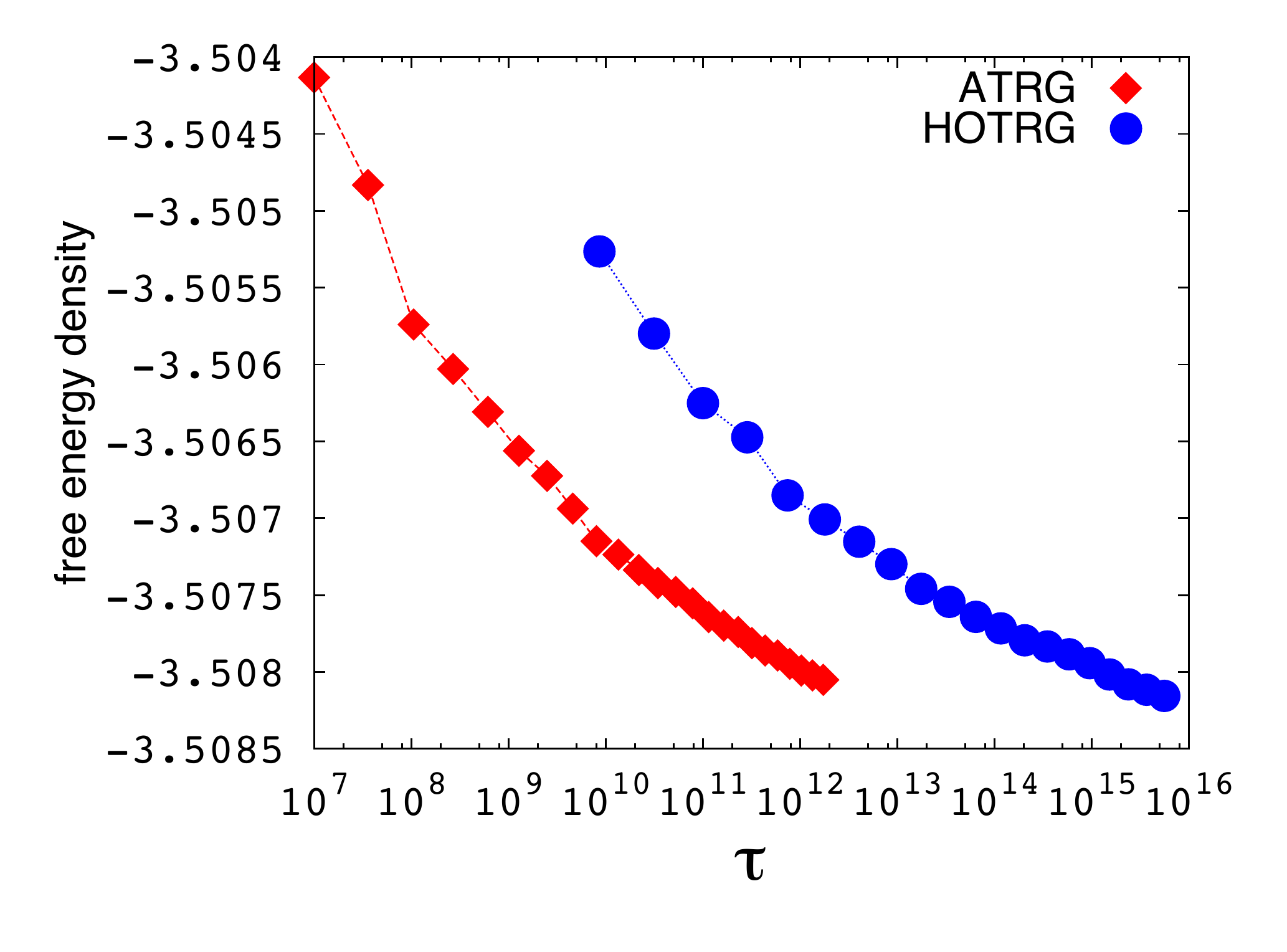}
  \vspace{20pt}
  \caption{
    Free energy density of the three-dimensional Ising model at $T=T_{\rm c}$ as a function of leading-order computation time $\tau$ [Eq.~\eqref{tau_3d}] calculated by HOTRG (blue circles) and ATRG (red diamonds). ATRG achieves much lower free energy than HOTRG with fixed computation time.
  }
  \label{Ising3D_2}
\end{figure}

Next, we move to the three-dimensional Ising model on a simple cubic lattice.
In the three-dimensional case, we compare ATRG with HOTRG.
In Fig.~\ref{Ising3D_2}, we show the free energy density as a function of $\tau$ at $T = T_{\rm c} = 4.5115$~\cite{XieCQZYX2012, DengB2003, Hasenbusch2010, KaupuzsMR2017}.
Here, we again define the leading-order computation time $\tau$ for three dimensions as 
\begin{align}
  \tau = \begin{cases}
    \chi^{7} & \text{for ATRG} \\
    \chi^{11} & \text{for HOTRG.}
  \end{cases}
  \label{tau_3d}
\end{align}
In the present case, we calculate up to $\chi=56$ ($\chi=27$) for ATRG (HOTRG).
In Fig.~\ref{Ising3D_2}, it is clearly demonstrated again that the ATRG gives the better (lower) free energy density than that of HOTRG for the same leading-order computation time $\tau$. We expect that the advantage of ATRG over HOTRG should be more pronounced in higher dimensions.

%{{{ RESULTS OF 3D ISING
%}}}

\section{Summary}
\label{sec:summary}

In the present paper, we proposed the ATRG method that can perform tensor renormalization operations with computation cost of $O(\chi^{2d+1})$ and memory footprint of $O(\chi^{d+1})$ for $d$-dimensional hyper cubic lattices.
The computation cost and the memory footprint of the proposed method are much lower than that of the conventional HOTRG, $O(\chi^{4d-1})$ and $O(\chi^{2d})$, respectively, which enables us to apply the tensor renormalization method in higher dimensions.
Unlike HOTRG, our algorithm involves the truncation of the bond dimension by using SVD when we swap the bonds of two tensors. 
Due to this additional approximation, the accuracy in the final result degrades compared with HOTRG.
However, this disadvantage is compensated by the drastic reduction of the computation cost from $O(\chi^{4d-1})$ to $O(\chi^{2d+1})$.
We confirmed that for two- and three-dimensional Ising models our method achieve higher accuracy than HOTRG with fixed leading-order computational time.
Since ATRG is a real space renormalization method similar to HOTRG, and preserves the lattice topology after the renormalization, it can be applied to various lattice systems in arbitrary dimensions

Finally, as we pointed out already, the partial SVD is the most expensive operation in ATRG.
The performance of the partial SVD is thus essential in ATRG, and development of more efficient and stable partial SVD algorithms is desired for future application of ATRG to large-scale complex lattice systems.

\section*{Acknowledgement}

The authors would like to thank Takuya Yamamoto and Hayate Nakano for careful reading of the manuscript and comments.
D.~A. is supported by the Japan Society for the Promotion of Science through the Program for Leading Graduate Schools (MERIT).
This work is partially supported by MEXT as “Exploratory Challenge on Post-K computer” (Frontiers of Basic Science: Challenging the Limits) and by JSPS KAKENHI No.~15K17701, 17K05564, and 19K03740. 

\bibliography{ATRG}

\begin{thebibliography}{37}%
\makeatletter
\providecommand \@ifxundefined [1]{%
 \@ifx{#1\undefined}
}%
\providecommand \@ifnum [1]{%
 \ifnum #1\expandafter \@firstoftwo
 \else \expandafter \@secondoftwo
 \fi
}%
\providecommand \@ifx [1]{%
 \ifx #1\expandafter \@firstoftwo
 \else \expandafter \@secondoftwo
 \fi
}%
\providecommand \natexlab [1]{#1}%
\providecommand \enquote  [1]{``#1''}%
\providecommand \bibnamefont  [1]{#1}%
\providecommand \bibfnamefont [1]{#1}%
\providecommand \citenamefont [1]{#1}%
\providecommand \href@noop [0]{\@secondoftwo}%
\providecommand \href [0]{\begingroup \@sanitize@url \@href}%
\providecommand \@href[1]{\@@startlink{#1}\@@href}%
\providecommand \@@href[1]{\endgroup#1\@@endlink}%
\providecommand \@sanitize@url [0]{\catcode `\\12\catcode `\$12\catcode
  `\&12\catcode `\#12\catcode `\^12\catcode `\_12\catcode `\%12\relax}%
\providecommand \@@startlink[1]{}%
\providecommand \@@endlink[0]{}%
\providecommand \url  [0]{\begingroup\@sanitize@url \@url }%
\providecommand \@url [1]{\endgroup\@href {#1}{\urlprefix }}%
\providecommand \urlprefix  [0]{URL }%
\providecommand \Eprint [0]{\href }%
\providecommand \doibase [0]{http://dx.doi.org/}%
\providecommand \selectlanguage [0]{\@gobble}%
\providecommand \bibinfo  [0]{\@secondoftwo}%
\providecommand \bibfield  [0]{\@secondoftwo}%
\providecommand \translation [1]{[#1]}%
\providecommand \BibitemOpen [0]{}%
\providecommand \bibitemStop [0]{}%
\providecommand \bibitemNoStop [0]{.\EOS\space}%
\providecommand \EOS [0]{\spacefactor3000\relax}%
\providecommand \BibitemShut  [1]{\csname bibitem#1\endcsname}%
\let\auto@bib@innerbib\@empty
%</preamble>
\bibitem [{\citenamefont {Levin}\ and\ \citenamefont
  {Nave}(2007)}]{LevinN2007}%
  \BibitemOpen
  \bibfield  {author} {\bibinfo {author} {\bibfnamefont {M.}~\bibnamefont
  {Levin}}\ and\ \bibinfo {author} {\bibfnamefont {C.~P.}\ \bibnamefont
  {Nave}},\ }\href {\doibase 10.1103/PhysRevLett.99.120601} {\bibfield
  {journal} {\bibinfo  {journal} {Phys. Rev. Lett.}\ }\textbf {\bibinfo
  {volume} {99}},\ \bibinfo {pages} {120601} (\bibinfo {year}
  {2007})}\BibitemShut {NoStop}%
\bibitem [{\citenamefont {Xie}\ \emph {et~al.}(2012)\citenamefont {Xie},
  \citenamefont {Chen}, \citenamefont {Qin}, \citenamefont {Zhu}, \citenamefont
  {Yang},\ and\ \citenamefont {Xiang}}]{XieCQZYX2012}%
  \BibitemOpen
  \bibfield  {author} {\bibinfo {author} {\bibfnamefont {Z.~Y.}\ \bibnamefont
  {Xie}}, \bibinfo {author} {\bibfnamefont {J.}~\bibnamefont {Chen}}, \bibinfo
  {author} {\bibfnamefont {M.~P.}\ \bibnamefont {Qin}}, \bibinfo {author}
  {\bibfnamefont {J.~W.}\ \bibnamefont {Zhu}}, \bibinfo {author} {\bibfnamefont
  {L.~P.}\ \bibnamefont {Yang}}, \ and\ \bibinfo {author} {\bibfnamefont
  {T.}~\bibnamefont {Xiang}},\ }\href {\doibase 10.1103/PhysRevB.86.045139}
  {\bibfield  {journal} {\bibinfo  {journal} {Phys. Rev. B}\ }\textbf {\bibinfo
  {volume} {86}},\ \bibinfo {pages} {045139} (\bibinfo {year}
  {2012})}\BibitemShut {NoStop}%
\bibitem [{\citenamefont {Gu}\ and\ \citenamefont {Wen}(2009)}]{GuW2009}%
  \BibitemOpen
  \bibfield  {author} {\bibinfo {author} {\bibfnamefont {Z.-C.}\ \bibnamefont
  {Gu}}\ and\ \bibinfo {author} {\bibfnamefont {X.-G.}\ \bibnamefont {Wen}},\
  }\href {\doibase 10.1103/PhysRevB.80.155131} {\bibfield  {journal} {\bibinfo
  {journal} {Phys. Rev. B}\ }\textbf {\bibinfo {volume} {80}},\ \bibinfo
  {pages} {155131} (\bibinfo {year} {2009})}\BibitemShut {NoStop}%
\bibitem [{\citenamefont {Xie}\ \emph {et~al.}(2009)\citenamefont {Xie},
  \citenamefont {Jiang}, \citenamefont {Chen}, \citenamefont {Weng},\ and\
  \citenamefont {Xiang}}]{XieJCWX2009}%
  \BibitemOpen
  \bibfield  {author} {\bibinfo {author} {\bibfnamefont {Z.~Y.}\ \bibnamefont
  {Xie}}, \bibinfo {author} {\bibfnamefont {H.~C.}\ \bibnamefont {Jiang}},
  \bibinfo {author} {\bibfnamefont {Q.~N.}\ \bibnamefont {Chen}}, \bibinfo
  {author} {\bibfnamefont {Z.~Y.}\ \bibnamefont {Weng}}, \ and\ \bibinfo
  {author} {\bibfnamefont {T.}~\bibnamefont {Xiang}},\ }\href {\doibase
  10.1103/PhysRevLett.103.160601} {\bibfield  {journal} {\bibinfo  {journal}
  {Phys. Rev. Lett.}\ }\textbf {\bibinfo {volume} {103}},\ \bibinfo {pages}
  {160601} (\bibinfo {year} {2009})}\BibitemShut {NoStop}%
\bibitem [{\citenamefont {Evenbly}\ and\ \citenamefont
  {Vidal}(2015)}]{EvenblyV2015}%
  \BibitemOpen
  \bibfield  {author} {\bibinfo {author} {\bibfnamefont {G.}~\bibnamefont
  {Evenbly}}\ and\ \bibinfo {author} {\bibfnamefont {G.}~\bibnamefont
  {Vidal}},\ }\href {\doibase 10.1103/PhysRevLett.115.180405} {\bibfield
  {journal} {\bibinfo  {journal} {Phys. Rev. Lett.}\ }\textbf {\bibinfo
  {volume} {115}},\ \bibinfo {pages} {180405} (\bibinfo {year}
  {2015})}\BibitemShut {NoStop}%
\bibitem [{\citenamefont {Evenbly}(2017)}]{Evenbly2017}%
  \BibitemOpen
  \bibfield  {author} {\bibinfo {author} {\bibfnamefont {G.}~\bibnamefont
  {Evenbly}},\ }\href {\doibase 10.1103/PhysRevB.95.045117} {\bibfield
  {journal} {\bibinfo  {journal} {Phys. Rev. B}\ }\textbf {\bibinfo {volume}
  {95}},\ \bibinfo {pages} {045117} (\bibinfo {year} {2017})}\BibitemShut
  {NoStop}%
\bibitem [{\citenamefont {Bal}\ \emph {et~al.}(2017)\citenamefont {Bal},
  \citenamefont {Mari\"en}, \citenamefont {Haegeman},\ and\ \citenamefont
  {Verstraete}}]{BalMHV2017}%
  \BibitemOpen
  \bibfield  {author} {\bibinfo {author} {\bibfnamefont {M.}~\bibnamefont
  {Bal}}, \bibinfo {author} {\bibfnamefont {M.}~\bibnamefont {Mari\"en}},
  \bibinfo {author} {\bibfnamefont {J.}~\bibnamefont {Haegeman}}, \ and\
  \bibinfo {author} {\bibfnamefont {F.}~\bibnamefont {Verstraete}},\ }\href
  {\doibase 10.1103/PhysRevLett.118.250602} {\bibfield  {journal} {\bibinfo
  {journal} {Phys. Rev. Lett.}\ }\textbf {\bibinfo {volume} {118}},\ \bibinfo
  {pages} {250602} (\bibinfo {year} {2017})}\BibitemShut {NoStop}%
\bibitem [{\citenamefont {Yang}\ \emph {et~al.}(2017)\citenamefont {Yang},
  \citenamefont {Gu},\ and\ \citenamefont {Wen}}]{YangGW2017}%
  \BibitemOpen
  \bibfield  {author} {\bibinfo {author} {\bibfnamefont {S.}~\bibnamefont
  {Yang}}, \bibinfo {author} {\bibfnamefont {Z.-C.}\ \bibnamefont {Gu}}, \ and\
  \bibinfo {author} {\bibfnamefont {X.-G.}\ \bibnamefont {Wen}},\ }\href
  {\doibase 10.1103/PhysRevLett.118.110504} {\bibfield  {journal} {\bibinfo
  {journal} {Phys. Rev. Lett.}\ }\textbf {\bibinfo {volume} {118}},\ \bibinfo
  {pages} {110504} (\bibinfo {year} {2017})}\BibitemShut {NoStop}%
\bibitem [{\citenamefont {Hauru}\ \emph {et~al.}(2018)\citenamefont {Hauru},
  \citenamefont {Delcamp},\ and\ \citenamefont {Mizera}}]{HauruDM2018}%
  \BibitemOpen
  \bibfield  {author} {\bibinfo {author} {\bibfnamefont {M.}~\bibnamefont
  {Hauru}}, \bibinfo {author} {\bibfnamefont {C.}~\bibnamefont {Delcamp}}, \
  and\ \bibinfo {author} {\bibfnamefont {S.}~\bibnamefont {Mizera}},\ }\href
  {\doibase 10.1103/PhysRevB.97.045111} {\bibfield  {journal} {\bibinfo
  {journal} {Phys. Rev. B}\ }\textbf {\bibinfo {volume} {97}},\ \bibinfo
  {pages} {045111} (\bibinfo {year} {2018})}\BibitemShut {NoStop}%
\bibitem [{\citenamefont {Harada}(2018)}]{Harada2018}%
  \BibitemOpen
  \bibfield  {author} {\bibinfo {author} {\bibfnamefont {K.}~\bibnamefont
  {Harada}},\ }\href {\doibase 10.1103/PhysRevB.97.045124} {\bibfield
  {journal} {\bibinfo  {journal} {Phys. Rev. B}\ }\textbf {\bibinfo {volume}
  {97}},\ \bibinfo {pages} {045124} (\bibinfo {year} {2018})}\BibitemShut
  {NoStop}%
\bibitem [{\citenamefont {Morita}\ \emph {et~al.}(2018)\citenamefont {Morita},
  \citenamefont {Igarashi}, \citenamefont {Zhao},\ and\ \citenamefont
  {Kawashima}}]{MoritaIZK2018}%
  \BibitemOpen
  \bibfield  {author} {\bibinfo {author} {\bibfnamefont {S.}~\bibnamefont
  {Morita}}, \bibinfo {author} {\bibfnamefont {R.}~\bibnamefont {Igarashi}},
  \bibinfo {author} {\bibfnamefont {H.-H.}\ \bibnamefont {Zhao}}, \ and\
  \bibinfo {author} {\bibfnamefont {N.}~\bibnamefont {Kawashima}},\ }\href
  {\doibase 10.1103/PhysRevE.97.033310} {\bibfield  {journal} {\bibinfo
  {journal} {Phys. Rev. E}\ }\textbf {\bibinfo {volume} {97}},\ \bibinfo
  {pages} {033310} (\bibinfo {year} {2018})}\BibitemShut {NoStop}%
\bibitem [{\citenamefont {Wang}\ \emph
  {et~al.}(2014{\natexlab{a}})\citenamefont {Wang}, \citenamefont {Qin},\ and\
  \citenamefont {Zhou}}]{WangQZ2014}%
  \BibitemOpen
  \bibfield  {author} {\bibinfo {author} {\bibfnamefont {C.}~\bibnamefont
  {Wang}}, \bibinfo {author} {\bibfnamefont {S.-M.}\ \bibnamefont {Qin}}, \
  and\ \bibinfo {author} {\bibfnamefont {H.-J.}\ \bibnamefont {Zhou}},\ }\href
  {\doibase 10.1103/PhysRevB.90.174201} {\bibfield  {journal} {\bibinfo
  {journal} {Phys. Rev. B}\ }\textbf {\bibinfo {volume} {90}},\ \bibinfo
  {pages} {174201} (\bibinfo {year} {2014}{\natexlab{a}})}\BibitemShut
  {NoStop}%
\bibitem [{\citenamefont {Zhao}\ \emph {et~al.}(2016)\citenamefont {Zhao},
  \citenamefont {Xie}, \citenamefont {Xiang},\ and\ \citenamefont
  {Imada}}]{ZhaoXXI2016}%
  \BibitemOpen
  \bibfield  {author} {\bibinfo {author} {\bibfnamefont {H.-H.}\ \bibnamefont
  {Zhao}}, \bibinfo {author} {\bibfnamefont {Z.-Y.}\ \bibnamefont {Xie}},
  \bibinfo {author} {\bibfnamefont {T.}~\bibnamefont {Xiang}}, \ and\ \bibinfo
  {author} {\bibfnamefont {M.}~\bibnamefont {Imada}},\ }\href {\doibase
  10.1103/PhysRevB.93.125115} {\bibfield  {journal} {\bibinfo  {journal} {Phys.
  Rev. B}\ }\textbf {\bibinfo {volume} {93}},\ \bibinfo {pages} {125115}
  (\bibinfo {year} {2016})}\BibitemShut {NoStop}%
\bibitem [{\citenamefont {Evenbly}(2018)}]{Evenbly2018}%
  \BibitemOpen
  \bibfield  {author} {\bibinfo {author} {\bibfnamefont {G.}~\bibnamefont
  {Evenbly}},\ }\href {\doibase 10.1103/PhysRevB.98.085155} {\bibfield
  {journal} {\bibinfo  {journal} {Phys. Rev. B}\ }\textbf {\bibinfo {volume}
  {98}},\ \bibinfo {pages} {085155} (\bibinfo {year} {2018})}\BibitemShut
  {NoStop}%
\bibitem [{\citenamefont {Ferris}(2013)}]{Ferris2013}%
  \BibitemOpen
  \bibfield  {author} {\bibinfo {author} {\bibfnamefont {A.~J.}\ \bibnamefont
  {Ferris}},\ }\href {\doibase 10.1103/PhysRevB.87.125139} {\bibfield
  {journal} {\bibinfo  {journal} {Phys. Rev. B}\ }\textbf {\bibinfo {volume}
  {87}},\ \bibinfo {pages} {125139} (\bibinfo {year} {2013})}\BibitemShut
  {NoStop}%
\bibitem [{\citenamefont {Li}\ \emph {et~al.}(2010)\citenamefont {Li},
  \citenamefont {Gong}, \citenamefont {Zhao}, \citenamefont {Ran},
  \citenamefont {Gao},\ and\ \citenamefont {Su}}]{LiGZRGS2010}%
  \BibitemOpen
  \bibfield  {author} {\bibinfo {author} {\bibfnamefont {W.}~\bibnamefont
  {Li}}, \bibinfo {author} {\bibfnamefont {S.-S.}\ \bibnamefont {Gong}},
  \bibinfo {author} {\bibfnamefont {Y.}~\bibnamefont {Zhao}}, \bibinfo {author}
  {\bibfnamefont {S.-J.}\ \bibnamefont {Ran}}, \bibinfo {author} {\bibfnamefont
  {S.}~\bibnamefont {Gao}}, \ and\ \bibinfo {author} {\bibfnamefont
  {G.}~\bibnamefont {Su}},\ }\href {\doibase 10.1103/PhysRevB.82.134434}
  {\bibfield  {journal} {\bibinfo  {journal} {Phys. Rev. B}\ }\textbf {\bibinfo
  {volume} {82}},\ \bibinfo {pages} {134434} (\bibinfo {year}
  {2010})}\BibitemShut {NoStop}%
\bibitem [{\citenamefont {Zhao}\ \emph {et~al.}(2010)\citenamefont {Zhao},
  \citenamefont {Xie}, \citenamefont {Chen}, \citenamefont {Wei}, \citenamefont
  {Cai},\ and\ \citenamefont {Xiang}}]{ZhaoXCWCX2010}%
  \BibitemOpen
  \bibfield  {author} {\bibinfo {author} {\bibfnamefont {H.~H.}\ \bibnamefont
  {Zhao}}, \bibinfo {author} {\bibfnamefont {Z.~Y.}\ \bibnamefont {Xie}},
  \bibinfo {author} {\bibfnamefont {Q.~N.}\ \bibnamefont {Chen}}, \bibinfo
  {author} {\bibfnamefont {Z.~C.}\ \bibnamefont {Wei}}, \bibinfo {author}
  {\bibfnamefont {J.~W.}\ \bibnamefont {Cai}}, \ and\ \bibinfo {author}
  {\bibfnamefont {T.}~\bibnamefont {Xiang}},\ }\href {\doibase
  10.1103/PhysRevB.81.174411} {\bibfield  {journal} {\bibinfo  {journal} {Phys.
  Rev. B}\ }\textbf {\bibinfo {volume} {81}},\ \bibinfo {pages} {174411}
  (\bibinfo {year} {2010})}\BibitemShut {NoStop}%
\bibitem [{\citenamefont {Chen}\ \emph {et~al.}(2011)\citenamefont {Chen},
  \citenamefont {Qin}, \citenamefont {Chen}, \citenamefont {Wei}, \citenamefont
  {Zhao}, \citenamefont {Normand},\ and\ \citenamefont
  {Xiang}}]{ChenQCWZNX2011}%
  \BibitemOpen
  \bibfield  {author} {\bibinfo {author} {\bibfnamefont {Q.~N.}\ \bibnamefont
  {Chen}}, \bibinfo {author} {\bibfnamefont {M.~P.}\ \bibnamefont {Qin}},
  \bibinfo {author} {\bibfnamefont {J.}~\bibnamefont {Chen}}, \bibinfo {author}
  {\bibfnamefont {Z.~C.}\ \bibnamefont {Wei}}, \bibinfo {author} {\bibfnamefont
  {H.~H.}\ \bibnamefont {Zhao}}, \bibinfo {author} {\bibfnamefont
  {B.}~\bibnamefont {Normand}}, \ and\ \bibinfo {author} {\bibfnamefont
  {T.}~\bibnamefont {Xiang}},\ }\href {\doibase 10.1103/PhysRevLett.107.165701}
  {\bibfield  {journal} {\bibinfo  {journal} {Phys. Rev. Lett.}\ }\textbf
  {\bibinfo {volume} {107}},\ \bibinfo {pages} {165701} (\bibinfo {year}
  {2011})}\BibitemShut {NoStop}%
\bibitem [{\citenamefont {Dittrich}\ and\ \citenamefont
  {Eckert}(2012)}]{DittrichE2012}%
  \BibitemOpen
  \bibfield  {author} {\bibinfo {author} {\bibfnamefont {B.}~\bibnamefont
  {Dittrich}}\ and\ \bibinfo {author} {\bibfnamefont {F.~C.}\ \bibnamefont
  {Eckert}},\ }\href {http://stacks.iop.org/1742-6596/360/i=1/a=012004}
  {\bibfield  {journal} {\bibinfo  {journal} {J. Phys.: Conf. Ser.}\ }\textbf
  {\bibinfo {volume} {360}},\ \bibinfo {pages} {012004} (\bibinfo {year}
  {2012})}\BibitemShut {NoStop}%
\bibitem [{\citenamefont {Jiang}\ \emph {et~al.}(2008)\citenamefont {Jiang},
  \citenamefont {Weng},\ and\ \citenamefont {Xiang}}]{JiangWX2008}%
  \BibitemOpen
  \bibfield  {author} {\bibinfo {author} {\bibfnamefont {H.~C.}\ \bibnamefont
  {Jiang}}, \bibinfo {author} {\bibfnamefont {Z.~Y.}\ \bibnamefont {Weng}}, \
  and\ \bibinfo {author} {\bibfnamefont {T.}~\bibnamefont {Xiang}},\ }\href
  {\doibase 10.1103/PhysRevLett.101.090603} {\bibfield  {journal} {\bibinfo
  {journal} {Phys. Rev. Lett.}\ }\textbf {\bibinfo {volume} {101}},\ \bibinfo
  {pages} {090603} (\bibinfo {year} {2008})}\BibitemShut {NoStop}%
\bibitem [{\citenamefont {Evenbly}\ and\ \citenamefont
  {Vidal}(2016)}]{EvenblyV2016}%
  \BibitemOpen
  \bibfield  {author} {\bibinfo {author} {\bibfnamefont {G.}~\bibnamefont
  {Evenbly}}\ and\ \bibinfo {author} {\bibfnamefont {G.}~\bibnamefont
  {Vidal}},\ }\href {\doibase 10.1103/PhysRevLett.116.040401} {\bibfield
  {journal} {\bibinfo  {journal} {Phys. Rev. Lett.}\ }\textbf {\bibinfo
  {volume} {116}},\ \bibinfo {pages} {040401} (\bibinfo {year}
  {2016})}\BibitemShut {NoStop}%
\bibitem [{\citenamefont {Meurice}(2013)}]{Meurice2013}%
  \BibitemOpen
  \bibfield  {author} {\bibinfo {author} {\bibfnamefont {Y.}~\bibnamefont
  {Meurice}},\ }\href {\doibase 10.1103/PhysRevB.87.064422} {\bibfield
  {journal} {\bibinfo  {journal} {Phys. Rev. B}\ }\textbf {\bibinfo {volume}
  {87}},\ \bibinfo {pages} {064422} (\bibinfo {year} {2013})}\BibitemShut
  {NoStop}%
\bibitem [{\citenamefont {Wang}\ \emph
  {et~al.}(2014{\natexlab{b}})\citenamefont {Wang}, \citenamefont {Xie},
  \citenamefont {Chen}, \citenamefont {Bruce},\ and\ \citenamefont
  {Xiang}}]{WangXCBX2014}%
  \BibitemOpen
  \bibfield  {author} {\bibinfo {author} {\bibfnamefont {S.}~\bibnamefont
  {Wang}}, \bibinfo {author} {\bibfnamefont {Z.-Y.}\ \bibnamefont {Xie}},
  \bibinfo {author} {\bibfnamefont {J.}~\bibnamefont {Chen}}, \bibinfo {author}
  {\bibfnamefont {N.}~\bibnamefont {Bruce}}, \ and\ \bibinfo {author}
  {\bibfnamefont {T.}~\bibnamefont {Xiang}},\ }\href
  {http://stacks.iop.org/0256-307X/31/i=7/a=070503} {\bibfield  {journal}
  {\bibinfo  {journal} {Chin. Phys. Lett.}\ }\textbf {\bibinfo {volume} {31}},\
  \bibinfo {pages} {070503} (\bibinfo {year} {2014}{\natexlab{b}})}\BibitemShut
  {NoStop}%
\bibitem [{\citenamefont {Yu}\ \emph {et~al.}(2014)\citenamefont {Yu},
  \citenamefont {Xie}, \citenamefont {Meurice}, \citenamefont {Liu},
  \citenamefont {Denbleyker}, \citenamefont {Zou}, \citenamefont {Qin},
  \citenamefont {Chen},\ and\ \citenamefont {Xiang}}]{YuXMLDZQCX2014}%
  \BibitemOpen
  \bibfield  {author} {\bibinfo {author} {\bibfnamefont {J.~F.}\ \bibnamefont
  {Yu}}, \bibinfo {author} {\bibfnamefont {Z.~Y.}\ \bibnamefont {Xie}},
  \bibinfo {author} {\bibfnamefont {Y.}~\bibnamefont {Meurice}}, \bibinfo
  {author} {\bibfnamefont {Y.}~\bibnamefont {Liu}}, \bibinfo {author}
  {\bibfnamefont {A.}~\bibnamefont {Denbleyker}}, \bibinfo {author}
  {\bibfnamefont {H.}~\bibnamefont {Zou}}, \bibinfo {author} {\bibfnamefont
  {M.~P.}\ \bibnamefont {Qin}}, \bibinfo {author} {\bibfnamefont
  {J.}~\bibnamefont {Chen}}, \ and\ \bibinfo {author} {\bibfnamefont
  {T.}~\bibnamefont {Xiang}},\ }\href {\doibase 10.1103/PhysRevE.89.013308}
  {\bibfield  {journal} {\bibinfo  {journal} {Phys. Rev. E}\ }\textbf {\bibinfo
  {volume} {89}},\ \bibinfo {pages} {013308} (\bibinfo {year}
  {2014})}\BibitemShut {NoStop}%
\bibitem [{\citenamefont {Ueda}\ \emph {et~al.}(2014)\citenamefont {Ueda},
  \citenamefont {Okunishi},\ and\ \citenamefont {Nishino}}]{UedaON2014}%
  \BibitemOpen
  \bibfield  {author} {\bibinfo {author} {\bibfnamefont {H.}~\bibnamefont
  {Ueda}}, \bibinfo {author} {\bibfnamefont {K.}~\bibnamefont {Okunishi}}, \
  and\ \bibinfo {author} {\bibfnamefont {T.}~\bibnamefont {Nishino}},\ }\href
  {\doibase 10.1103/PhysRevB.89.075116} {\bibfield  {journal} {\bibinfo
  {journal} {Phys. Rev. B}\ }\textbf {\bibinfo {volume} {89}},\ \bibinfo
  {pages} {075116} (\bibinfo {year} {2014})}\BibitemShut {NoStop}%
\bibitem [{\citenamefont {Genzor}\ \emph {et~al.}(2016)\citenamefont {Genzor},
  \citenamefont {Gendiar},\ and\ \citenamefont {Nishino}}]{GenzorGN2016}%
  \BibitemOpen
  \bibfield  {author} {\bibinfo {author} {\bibfnamefont {J.}~\bibnamefont
  {Genzor}}, \bibinfo {author} {\bibfnamefont {A.}~\bibnamefont {Gendiar}}, \
  and\ \bibinfo {author} {\bibfnamefont {T.}~\bibnamefont {Nishino}},\ }\href
  {\doibase 10.1103/PhysRevE.93.012141} {\bibfield  {journal} {\bibinfo
  {journal} {Phys. Rev. E}\ }\textbf {\bibinfo {volume} {93}},\ \bibinfo
  {pages} {012141} (\bibinfo {year} {2016})}\BibitemShut {NoStop}%
\bibitem [{\citenamefont {Yang}\ \emph {et~al.}(2016)\citenamefont {Yang},
  \citenamefont {Liu}, \citenamefont {Zou}, \citenamefont {Xie},\ and\
  \citenamefont {Meurice}}]{YangLZXM2016}%
  \BibitemOpen
  \bibfield  {author} {\bibinfo {author} {\bibfnamefont {L.-P.}\ \bibnamefont
  {Yang}}, \bibinfo {author} {\bibfnamefont {Y.}~\bibnamefont {Liu}}, \bibinfo
  {author} {\bibfnamefont {H.}~\bibnamefont {Zou}}, \bibinfo {author}
  {\bibfnamefont {Z.~Y.}\ \bibnamefont {Xie}}, \ and\ \bibinfo {author}
  {\bibfnamefont {Y.}~\bibnamefont {Meurice}},\ }\href {\doibase
  10.1103/PhysRevE.93.012138} {\bibfield  {journal} {\bibinfo  {journal} {Phys.
  Rev. E}\ }\textbf {\bibinfo {volume} {93}},\ \bibinfo {pages} {012138}
  (\bibinfo {year} {2016})}\BibitemShut {NoStop}%
\bibitem [{\citenamefont {Kawauchi}\ and\ \citenamefont
  {Takeda}(2016)}]{KawauchiT2016}%
  \BibitemOpen
  \bibfield  {author} {\bibinfo {author} {\bibfnamefont {H.}~\bibnamefont
  {Kawauchi}}\ and\ \bibinfo {author} {\bibfnamefont {S.}~\bibnamefont
  {Takeda}},\ }\href {\doibase 10.1103/PhysRevD.93.114503} {\bibfield
  {journal} {\bibinfo  {journal} {Phys. Rev. D}\ }\textbf {\bibinfo {volume}
  {93}},\ \bibinfo {pages} {114503} (\bibinfo {year} {2016})}\BibitemShut
  {NoStop}%
\bibitem [{\citenamefont {Sakai}\ \emph {et~al.}(2017)\citenamefont {Sakai},
  \citenamefont {Takeda},\ and\ \citenamefont {Yoshimura}}]{SakaiTY2017}%
  \BibitemOpen
  \bibfield  {author} {\bibinfo {author} {\bibfnamefont {R.}~\bibnamefont
  {Sakai}}, \bibinfo {author} {\bibfnamefont {S.}~\bibnamefont {Takeda}}, \
  and\ \bibinfo {author} {\bibfnamefont {Y.}~\bibnamefont {Yoshimura}},\
  }\href@noop {} {\bibfield  {journal} {\bibinfo  {journal} {Prog. Theor. Exp.
  Phys.}\ }\textbf {\bibinfo {volume} {2017}},\ \bibinfo {pages} {063B07}
  (\bibinfo {year} {2017})}\BibitemShut {NoStop}%
\bibitem [{\citenamefont {Yoshimura}\ \emph {et~al.}(2018)\citenamefont
  {Yoshimura}, \citenamefont {Kuramashi}, \citenamefont {Nakamura},
  \citenamefont {Takeda},\ and\ \citenamefont {Sakai}}]{YoshimuraKNTS2018}%
  \BibitemOpen
  \bibfield  {author} {\bibinfo {author} {\bibfnamefont {Y.}~\bibnamefont
  {Yoshimura}}, \bibinfo {author} {\bibfnamefont {Y.}~\bibnamefont
  {Kuramashi}}, \bibinfo {author} {\bibfnamefont {Y.}~\bibnamefont {Nakamura}},
  \bibinfo {author} {\bibfnamefont {S.}~\bibnamefont {Takeda}}, \ and\ \bibinfo
  {author} {\bibfnamefont {R.}~\bibnamefont {Sakai}},\ }\href {\doibase
  10.1103/PhysRevD.97.054511} {\bibfield  {journal} {\bibinfo  {journal} {Phys.
  Rev. D}\ }\textbf {\bibinfo {volume} {97}},\ \bibinfo {pages} {054511}
  (\bibinfo {year} {2018})}\BibitemShut {NoStop}%
\bibitem [{\citenamefont {Akiyama}\ \emph {et~al.}()\citenamefont {Akiyama},
  \citenamefont {Kuramashi}, \citenamefont {Yamashita},\ and\ \citenamefont
  {Yoshimura}}]{AkiyamaKYY2019}%
  \BibitemOpen
  \bibfield  {author} {\bibinfo {author} {\bibfnamefont {S.}~\bibnamefont
  {Akiyama}}, \bibinfo {author} {\bibfnamefont {Y.}~\bibnamefont {Kuramashi}},
  \bibinfo {author} {\bibfnamefont {T.}~\bibnamefont {Yamashita}}, \ and\
  \bibinfo {author} {\bibfnamefont {Y.}~\bibnamefont {Yoshimura}},\ }\href@noop
  {} {\ }\Eprint {http://arxiv.org/abs/hep-lat/1906.06060}
  {arXiv:hep-lat/1906.06060} \BibitemShut {NoStop}%
\bibitem [{\citenamefont {Nakamura}\ \emph {et~al.}(2019)\citenamefont
  {Nakamura}, \citenamefont {Oba},\ and\ \citenamefont
  {Takeda}}]{NakamuraOT2019}%
  \BibitemOpen
  \bibfield  {author} {\bibinfo {author} {\bibfnamefont {Y.}~\bibnamefont
  {Nakamura}}, \bibinfo {author} {\bibfnamefont {H.}~\bibnamefont {Oba}}, \
  and\ \bibinfo {author} {\bibfnamefont {S.}~\bibnamefont {Takeda}},\ }\href
  {\doibase 10.1103/PhysRevB.99.155101} {\bibfield  {journal} {\bibinfo
  {journal} {Phys. Rev. B}\ }\textbf {\bibinfo {volume} {99}},\ \bibinfo
  {pages} {155101} (\bibinfo {year} {2019})}\BibitemShut {NoStop}%
\bibitem [{\citenamefont {Corboz}\ \emph {et~al.}(2014)\citenamefont {Corboz},
  \citenamefont {Rice},\ and\ \citenamefont {Troyer}}]{CorbozRT2014}%
  \BibitemOpen
  \bibfield  {author} {\bibinfo {author} {\bibfnamefont {P.}~\bibnamefont
  {Corboz}}, \bibinfo {author} {\bibfnamefont {T.~M.}\ \bibnamefont {Rice}}, \
  and\ \bibinfo {author} {\bibfnamefont {M.}~\bibnamefont {Troyer}},\ }\href
  {\doibase 10.1103/PhysRevLett.113.046402} {\bibfield  {journal} {\bibinfo
  {journal} {Phys. Rev. Lett.}\ }\textbf {\bibinfo {volume} {113}},\ \bibinfo
  {pages} {046402} (\bibinfo {year} {2014})}\BibitemShut {NoStop}%
\bibitem [{\citenamefont {Onsager}(1944)}]{Onsager1944}%
  \BibitemOpen
  \bibfield  {author} {\bibinfo {author} {\bibfnamefont {L.}~\bibnamefont
  {Onsager}},\ }\href {\doibase 10.1103/PhysRev.65.117} {\bibfield  {journal}
  {\bibinfo  {journal} {Phys. Rev.}\ }\textbf {\bibinfo {volume} {65}},\
  \bibinfo {pages} {117} (\bibinfo {year} {1944})}\BibitemShut {NoStop}%
\bibitem [{\citenamefont {Deng}\ and\ \citenamefont
  {Bl\"ote}(2003)}]{DengB2003}%
  \BibitemOpen
  \bibfield  {author} {\bibinfo {author} {\bibfnamefont {Y.}~\bibnamefont
  {Deng}}\ and\ \bibinfo {author} {\bibfnamefont {H.~W.~J.}\ \bibnamefont
  {Bl\"ote}},\ }\href {\doibase 10.1103/PhysRevE.68.036125} {\bibfield
  {journal} {\bibinfo  {journal} {Phys. Rev. E}\ }\textbf {\bibinfo {volume}
  {68}},\ \bibinfo {pages} {036125} (\bibinfo {year} {2003})}\BibitemShut
  {NoStop}%
\bibitem [{\citenamefont {Hasenbusch}(2010)}]{Hasenbusch2010}%
  \BibitemOpen
  \bibfield  {author} {\bibinfo {author} {\bibfnamefont {M.}~\bibnamefont
  {Hasenbusch}},\ }\href {\doibase 10.1103/PhysRevB.82.174433} {\bibfield
  {journal} {\bibinfo  {journal} {Phys. Rev. B}\ }\textbf {\bibinfo {volume}
  {82}},\ \bibinfo {pages} {174433} (\bibinfo {year} {2010})}\BibitemShut
  {NoStop}%
\bibitem [{\citenamefont {Kaupu\v{z}s}\ \emph {et~al.}(2017)\citenamefont
  {Kaupu\v{z}s}, \citenamefont {Melnik},\ and\ \citenamefont
  {Rim\v{s}\=ans}}]{KaupuzsMR2017}%
  \BibitemOpen
  \bibfield  {author} {\bibinfo {author} {\bibfnamefont {J.}~\bibnamefont
  {Kaupu\v{z}s}}, \bibinfo {author} {\bibfnamefont {R.~V.~N.}\ \bibnamefont
  {Melnik}}, \ and\ \bibinfo {author} {\bibfnamefont {J.}~\bibnamefont
  {Rim\v{s}\=ans}},\ }\href@noop {} {\bibfield  {journal} {\bibinfo  {journal}
  {Int. J. Mod. Phys. C}\ }\textbf {\bibinfo {volume} {28}},\ \bibinfo {pages}
  {1750044} (\bibinfo {year} {2017})}\BibitemShut {NoStop}%
\end{thebibliography}%

\end{document}